\documentclass[letter]{aa}  

\usepackage{graphicx}
\usepackage{txfonts}
\usepackage{placeins} 
\usepackage{dblfloatfix}
\usepackage[most]{tcolorbox}

\newcommand{\thCII}{$^{13}$C{\scriptsize\,II}}
\newcommand{\twCII}{$^{12}$C{\scriptsize\,II}}
\newcommand{\CII}{C{\scriptsize\,II}}

\newcommand{\HII}{H{\scriptsize\,II~}}

\begin{document} 

   \title{Space as a spectroscopic laboratory: High-resolution spectroscopy of the [$^{13}$C\,II] hyperfine structure with SOFIA/upGREAT}

   \author{S.~Kabanovic \inst{1,2} , V.~Ossenkopf-Okada \inst{1}, S.~Schlemmer\inst{1}, J.~Stutzki\inst{1}, N.~Schneider\inst{1},  U.\,U.~Graf\inst{1}, O.~Asvany\inst{1},  D.\,A.~Riechers\inst{1}, C.~Guevara\inst{3}, R.~Higgins\inst{1},  R.~Simon\inst{1}, Y.~Okada\inst{1}, M.~Mertens\inst{1, 4},  L.~Schneider\inst{1}, R.~Güsten\inst{4} and A.\,G.\,G.\,M.~Tielens\inst{5,6}
    }

   \institute{
   I. Physikalisches Institut, Universität zu Köln, Zülpicher Str. 77, 50937 Köln, Germany\\ \email{kabanovic@ph1.uni-koeln.de} \and
    Astronomisches Rechen-Institut, Zentrum für Astronomie der Universität Heidelberg, Mönchhofstraße 12-14, 69120 Heidelberg, Germany \and
    Instituto de Astronomía, Universidad Católica del Norte, Av. Angamos 0610, 1270398 Antofagasta, Chile\and 
    Max-Planck-Institut für Radioastronomie, Auf dem Hügel 69, 53121 Bonn, Germany\and 
    Department of Astronomy, University of Maryland, College Park, MD 20742, USA\and
    Leiden Observatory, Leiden University, PO Box 9513, 2300 RA Leiden, The Netherlands
    }
         
\date{Received 20 February 2026 / Accepted 4 May 2026}
\titlerunning{Astronomically determined [\thCII] hyperfine structure constants}
\authorrunning{S. Kabanovic}  

\abstract
{The [\twCII] emission at 158$\,{\rm \mu m}$ is a key cooling line of the interstellar medium and serves as a unique tracer of gas kinematics in spatially and spectrally resolved observations. However, its spectral profile is often modified by optical depth effects. The intrinsic line shape can be reconstructed via comparisons with emission from the less abundant singly ionized $^{13}$C isotope.

Due to the additional neutron spin, the [\thCII] emission splits into three hyperfine structure (hfs) transitions. Laboratory measurements have provided the [\thCII] centroid frequency and the frequency of the strongest hfs component ($F=2-1$), whereas the two weaker outer components ($F=1-0$ and $F=1-1$) have been inferred solely from quantum-mechanical calculations. Consequently, the magnetic-dipole hfs constants, from which the individual transition frequencies can be derived, have not been determined experimentally thus far.

The high spectral resolution of observations with the upgraded German Receiver for Astronomy at Terahertz Frequencies (upGREAT) on board SOFIA enabled the simultaneous detection of all three [\thCII] hfs transitions. Thus, we were able to determine, for the first time based on astronomical observations, the magnetic-dipole hyperfine structure constants, $A_{1/2}^{\rm hf} = 810.71(11)\,{\rm MHz}$ and $A_{3/2}^{\rm hf} = 162.18(5)\,{\rm MHz}$, of the [\thCII] $2s^2\,2p\,{}^{2}P^{\circ}$ ground term with unprecedented precision. Combining these with the previously measured [\thCII] centroid frequency in the laboratory, we derived the rest frequencies of all three hfs lines. Using the observed [\twCII] emission as a reference, we also determined the [\thCII] centroid frequency with improved precision from astronomical observations.

This work demonstrates that spectrally resolved astronomical observations can be used to constrain fundamental atomic properties. A high spectral resolution and signal-to-noise ratio enable a level of precision in determining hfs constants that rivals laboratory measurements. The approach is broadly applicable to other atomic and molecular transitions where laboratory data are difficult to obtain.} 
  

\maketitle

\section{Introduction}\label{sec:intro}

    The [\twCII] $158\,\mathrm{\mu m}$ fine-structure line (${^2P}_{3/2} - {^2P}_{1/2}$) is one of the most important cooling lines of the interstellar medium \citep{Tielens1985}. The emission originates largely from photodissociation regions \citep[PDRs,][]{ Hollenbach1999}, the interfaces between hot ionized \HII\ regions and cold molecular clouds. In addition, recent studies have shown that [\twCII] is a unique tracer of gas kinematics, enabling direct observations of the gas expansion around \HII\ regions \citep{Pabst2019, Pabst2020, Luisi2021, Tiwari2021, Beuther2022, Bonne2022, Faerber2025}. 
    
    However, the [\twCII] line can be strongly affected by optical depth effects \citep{Graf2012, Ossenkopf2013, Guevara2020, Guevara2024, Kabanovic2022} in dense, bright PDRs. In particular, absorption by cold gas along the line of sight (LOS) can alter the spectral profile, producing dips near the bulk velocity that can create the appearance of multiple velocity components. In order to disentangle optical depth effects from gas kinematics, the rarer isotope [\thCII] can be utilized. The low abundance of the $^{13}$C-isotope \citep[$\sim 70$ times less abundant in the solar neighborhood, ][]{Wilson1999} leads to optically thin transitions that simply integrate over the line-of-sight emission. 
    The coupling of the nuclear spin of $\rm ^{13}C^{+}$ ($I=1/2$), arising from its unpaired neutron, with the electron angular momentum $J$ splits the [$^{13}$\CII] ${}^2P_{3/2} - {}^2P_{1/2}$ transition into three hyperfine structure (hfs) lines: $F=2 - 1$, $F=1 - 0$, and $F=1 - 1$, where $F$ is the total angular momentum of the system.
    The corresponding hyperfine energy shifts for a level with quantum numbers~$(F,I,J)$ are
    \begin{equation}
        E^\mathrm{hf}_{F,I,J} =
        \frac{A^{\rm hf}_J}{2}\left[ F(F+1)-I(I+1)-J(J+1) \right]~.
        \label{eq:energy_hfs}
    \end{equation} 
    The magnetic-dipole hfs constants $A^{\rm hf}_J$ can be determined from the energy differences between these three transitions,
    \begin{align}
        \Delta E_{2-1} &= E^\mathrm{hf}_{F=2,\,I=1/2,\,J=3/2} - E^\mathrm{hf}_{F=1,\,I=1/2,\,J=3/2} = 2 A^{\rm hf}_{3/2} \nonumber,\\[4pt]
        \Delta E_{1-0} &= E^\mathrm{hf}_{F=1,\,I=1/2,\,J=1/2} - E^\mathrm{hf}_{F=0,\,I=1/2,\,J=1/2} = A^{\rm hf}_{1/2} .
        \label{eq:energy_to_hfs_const}
    \end{align} 
    A schematic illustration of the energy levels and an in-depth explanation of the notation used can be found in Appendix~\ref{sec:theory}. The low relative abundance of $^{13}$C$^+$ relative to $^{12}$C$^+$ results in weak lines, challenging the observational sensitivity requirements. The first detection of the $F=2 - 1$ component (in M~42) was reported by \citet{Boreiko1988}, using the Kuiper Airborne Observatory,  followed by the first detections of the second strongest [\thCII] $F=1-0$ line by \cite{Stacey1991b} and \cite{Boreiko1996}. The improved sensitivity of the upgraded German Receiver for Astronomy at Terahertz Frequencies \citep[upGREAT;][]{Heyminck2012, Risacher2016ITTST} on board the Stratospheric Observatory for Infrared Astronomy \citep[SOFIA;][]{Young2012} has enabled routine detections of [\thCII] lines alongside [\twCII] in Galactic sources \citep{Graf2012, Guevara2020, Kabanovic2022}. In addition, the first detection of [\thCII] emission outside the Milky Way was reported by \cite{Okada2019} in the Large Magellanic Cloud.
    
    Analysis of [\thCII] emission combined with [\twCII] emission allows us to derive the column density, optical depth, and excitation temperature of C$^+$ individually for each velocity component \citep{Kabanovic2022}. For this analysis, the hfs lines are shifted to a common velocity frame by correcting for their relative frequency offsets and then averaged, weighted by their relative intensities \citep{Guevara2020}, thereby increasing the signal-to-noise ratio (S/N) of the observations. To apply the frequency correction, it is essential to accurately determine the frequency shifts between the different transitions. This is especially important if the lines have only been marginally detected.

    Because of the highly reactive nature of C$^+$ and the dipole-forbidden character of the ${^2P}_{3/2} - {^2P}_{1/2}$ transition, observing this transition under laboratory conditions is challenging. Nevertheless, \cite{Cooksy1986} successfully determined the [\thCII] hfs transition frequencies using laser magnetic resonance combined with quantum-mechanical calculations. They experimentally derived the [\thCII] centroid and the combination of magnetic-dipole hfs constants, $(A_{1/2}^{\rm hf}-3A_{3/2}^{\rm hf})/4$, which determines the frequency shift between the centroid and the strongest $F=2-1$ line. The two remaining hfs transition lines, $F=1-0$ and $F=1-1$, were estimated using Hartree--Fock calculations from \cite{Schaefer_and_Klemm1970}. While the relative intensities of these transitions were initially reported incorrectly by \cite{Cooksy1986}, they were later corrected by \cite{Ossenkopf2013}. The resulting [\thCII] hfs frequencies, as summarized in Table~\ref{tab:freq_overview}, remain the only laboratory-based measurements and have long served as a reference for the astronomical community.

    In addition to astronomical observations, significant progress has also been made on the modeling side. \cite{Kramida_and_Haris2022} compiled all available data describing the energy levels of the C$^+$ isotopes, including, in particular, new multiconfiguration Dirac--Hartree--Fock computations from \cite{Joensson1996,Joensson2010}, which provide updated theoretical values of the hfs constants. Using the [\thCII] centroid frequency measured by \cite{Cooksy1986} as a reference, the accuracy of the three hyperfine transition frequencies can be improved through these more precise hfs constants. The updated theoretical [\thCII] frequencies, along with the corresponding calculation, are provided in Appendix~\ref{sec:theory}.     
\section{Astronomical observations}\label{sec:observations}
    
    \begin{figure}[tp]
        \centering
        \includegraphics[width=.99\columnwidth]{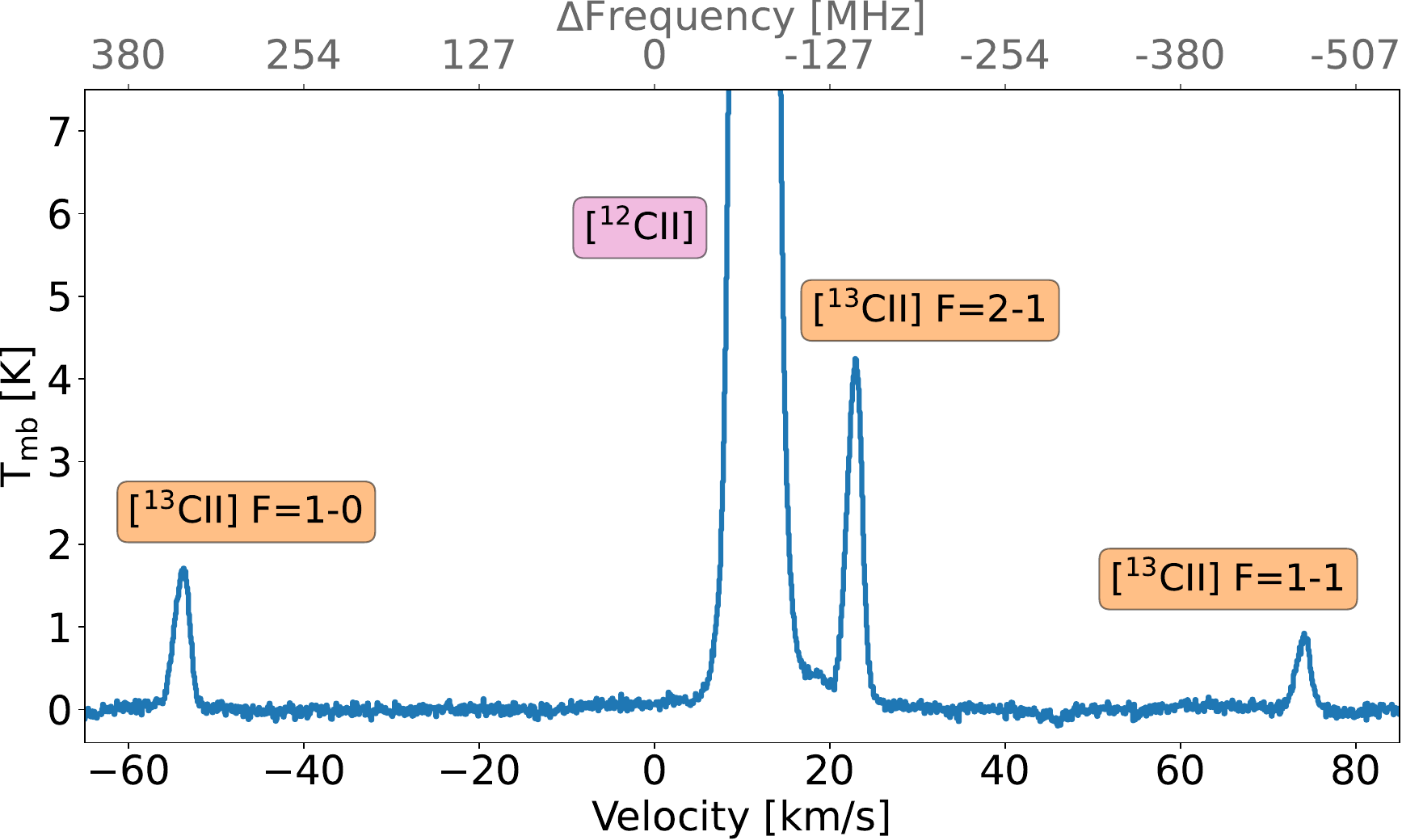}
        \caption{Spectrally resolved [\twCII] and [\thCII] emission. The spectrum shown was observed toward the emission peak of the bright rim in NGC~1977.}
            \label{fig:13cii_lines}
    \end{figure}
    The only direct simultaneous measurements of all three hfs lines have been obtained with astronomical observations. 
    An example spectrum with all three [\thCII] hfs lines together with the [\twCII] line is shown in Fig.~\ref{fig:13cii_lines}. The spectrum was observed in a single pointing long integration toward the peak of the bright rim of NGC~1977.
    Such an astronomical observation does not provide direct access to the intrinsic line frequencies, since the observed emission is Doppler shifted by the motion of the emitting source, preventing a unique separation of the intrinsic frequency and the source velocity. However, as all three [\thCII] lines are well detected, the frequency shifts between the hfs lines can be derived from these high spectral resolution data.
     
    To detect all three hfs lines, only bright sources with relatively narrow [\twCII] line widths are suitable. Therefore, we utilized the deep integrations of the [\twCII] and [\thCII] lines across Orion, including the Orion Bar, NGC~1977, and NGC~2024, taking advantage of the unprecedented sensitivity of upGREAT on board SOFIA. More information on the observation and data reduction is given in 
    Appendix~\ref{sec:data_reduction}.

\section{The velocity shifts of the [\thCII] lines}\label{sec:hfs_shift}
    
    In astronomical observations, it is common to transform the frequency axis into a velocity axis using the Doppler shift in non-relativistic approximation (conveniently resulting in a linear velocity scale), according to
    \begin{equation}
        v_{F'-F} = c\left(1-\frac{f_{F'-F}}{f_{\rm  [^{12}CII]}}\right),~ 
    \end{equation}
    \noindent
    with the [\twCII] rest frequency as reference frequency. Following astronomical convention, we adopted the velocity axis rather than the frequency axis. However, all results summarized in tables and figures are given both in frequency and in velocity offset relative to the [\twCII] line. With the above transformation, the [\twCII] line is Doppler shifted by the emitting cloud's systemic LOS velocity. 

    Since astronomical observations lack an absolute frequency reference, only the velocity differences between spectral lines can be determined directly from astronomical data. To this end, we perform a multi-Gaussian fit to the observed [\thCII] lines, assuming the emission to be optically thin and therefore representable as a superposition of Gaussian components. This fit provides access to the velocity differences between the three hfs lines and, through Eq.~(\ref{eq:energy_to_hfs_const}), to the hfs constants $A_{1/2}^{\rm hf}$ and $A_{3/2}^{\rm hf}$. An in-depth discussion of the fitting process and its results is presented in Appendix~\ref{sec:gauss_fit}. 

\begin{figure}[!htp]
        \centering
        \includegraphics[width=1.\columnwidth]{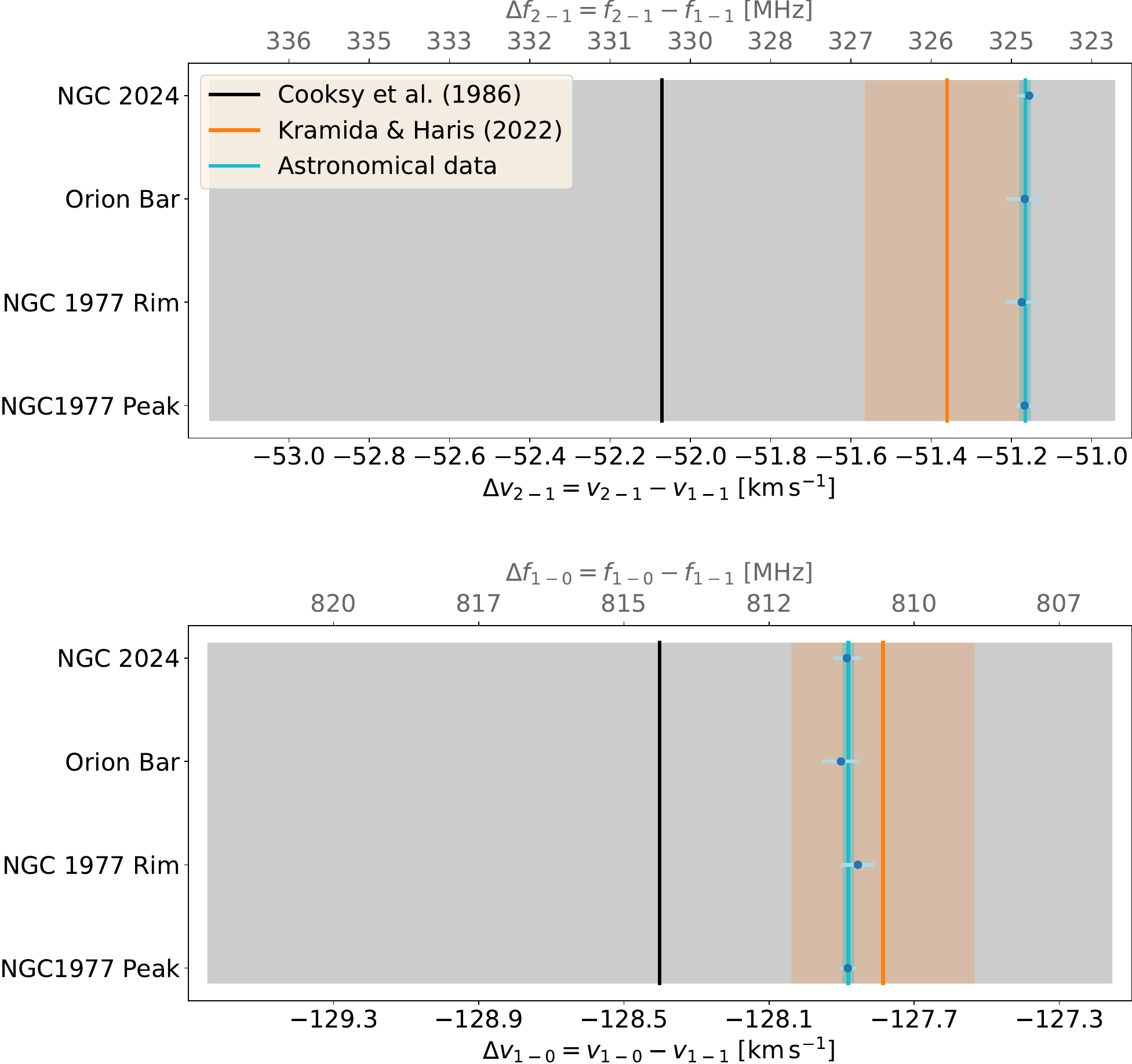}
    	\caption{Velocity and frequency differences between the [\thCII] hyperfine transitions. The upper panel shows the difference between the $F = 2-1$ and $F = 1-1$ transitions, and the lower panel between the $F = 1-0$ and $F = 1-1$ transitions. Black vertical lines indicate the differences determined from \cite{Cooksy1986}, with the gray shaded area denoting the corresponding uncertainty. Orange lines indicate the differences reported by \cite{Kramida_and_Haris2022}, derived from theoretical calculations by \cite{Joensson1996, Joensson2010}, with the orange shaded area denoting the corresponding uncertainty. Blue data points with error bars represent the differences determined from astronomical observations, while the cyan vertical line with shaded area indicates their mean and standard deviation, respectively.}
        \label{fig:13cii_hfs_shift}
    \end{figure}

    The frequency differences between the hyperfine states derived from the four astronomical observations are shown in Fig.~\ref{fig:13cii_hfs_shift} as blue data points with error bars. Since astronomical measurements provide direct access to the hfs transition lines, they yield the most direct determination of the hfs constants to date. The updated theoretical hfs constants reported by \cite{Kramida_and_Haris2022} are in close agreement with our observational results, with a small offset of $\sim 1\,{\rm MHz}$ that remains within their theoretical uncertainties. Earlier predictions by \cite{Cooksy1986} deviate more noticeably from the observed values, although their relatively large uncertainties still overlap with the astronomical results. Thanks to the high spectral resolution of our data, we constrain the hfs constants with an order of magnitude greater precision than previous numerical determinations, providing experimentally derived values for the first time. The constants obtained from the combined analysis of all four astronomical sources are listed in Table~\ref{tab:hfs_astro}.

    \begin{table}[!tb]
    \caption{[$^{13}$\CII] magnetic-dipole hfs constants derived from astronomical observations.}
    \label{tab:hfs_astro}
    \setlength{\tabcolsep}{3pt}
    \begin{tabular}{l c r r@{ $\pm$ }l r}
    \hline
    \hline
    Level & $F'-F$ & $\Delta f_{F'-F}^{(1)}$ & \multicolumn{2}{c}{$\Delta v_{F'-F}^{(2)}$} & $A_J^{\rm hf\,(3)}$ \\
    & & [MHz] & \multicolumn{2}{c}{[km$\,$s$^{-1}$]} & [MHz] \\
    \hline
    $2s^2\,2p\,{}^{2}P^{\circ}_{1/2}$ & $1-0$ & $810.71(11)$ & $-127.882$ & $0.017$ & $810.71(11)$ \\
    $2s^2\,2p\,{}^{2}P^{\circ}_{3/2}$ & $2-1$ & $324.37(9)$  & $-51.166$   & $0.015$ & $162.18(5)$  \\
    \hline
    \end{tabular}
    \tablefoot{(1) Frequency differences between the [\thCII] hfs lines derived from astronomical observations. The number in parentheses indicates the uncertainty in the last digits.
    (2) Corresponding velocity differences between the [\thCII] hfs lines.
    (3) [\thCII] hfs constants derived from astronomical observations.}
\end{table}

    In the second step, we combined the hfs constants derived in this work with the [\thCII] centroid frequency measured in the laboratory by \cite{Cooksy1986} to determine the absolute [\thCII] transition frequencies, following the calculations described in Appendix~\ref{sec:theory}. This yields [\thCII] transition frequencies derived purely from experimental data, listed in Table~\ref{tab:freq_astro}. Although the hfs constants are determined with exceptionally high accuracy, the uncertainty of the absolute rest frequencies is ultimately limited by the uncertainty of the reference frequency, which in this case corresponds to the laboratory-measured centroid frequency of \cite{Cooksy1986}. The corresponding velocity shifts with respect to the [\twCII] line are also provided; since these are derived from the rest frequencies, their uncertainties are likewise dominated by the uncertainty of the reference frequency.

     An alternative approach to determining the [\thCII] frequencies is to use the [\twCII] line as a reference by performing a simultaneous fit to the [\twCII] and [\thCII] transitions in astronomically observed spectra. The velocity shift between the [\twCII] and [\thCII] lines obtained from the fit, combined with the known [\twCII] rest frequency, yields the absolute [\thCII] frequencies. This approach is straightforward if both the [\twCII] and [\thCII] transitions are optically thin. However, the observed [\twCII] transitions are often affected by self- and foreground absorption, which distorts the line profile. This makes it unfeasible to determine the systemic velocity with the high accuracy required.

    By carefully examining each source, we find that only the single-pointing observation of the Orion Bar exhibits a sufficiently “intact” [\twCII] profile. Nevertheless, the [\twCII] emission from the Orion Bar is also optically thick \citep{Ossenkopf2013}, meaning that a simple Gaussian fit is not appropriate. Therefore, to properly model the [\twCII] emission, we fit the radiative transfer equation assuming that the line emission can be well described by a single excitation temperature. An in-depth discussion of the radiative transfer model and its application to the observed spectrum is provided in Appendix~\ref{sec:single_layer_model}.

    The model fit yields a velocity shift between the [\twCII] line and the [\thCII] $F=2-1$ transition of $11.357 \pm 0.009$~km~s$^{-1}$. Combining this result with the known [\twCII] rest frequency \citep{Cooksy1986}, we can determine the transition frequencies of the individual [\thCII] hyperfine components. The resulting frequencies are summarized in Table~\ref{tab:freq_astro}. As before, the uncertainty of the derived frequencies is dominated by the uncertainty of the reference line; however, since the [\twCII] line frequency is known more accurately than the [\thCII] centroid, the resulting [\thCII] frequencies have correspondingly lower uncertainties. Moreover, since this approach provides direct access to the velocity difference between the [\twCII] line and the [\thCII] hfs transitions, we can derive the velocity offsets, which are stated in Table~\ref{tab:freq_astro}, at the precision afforded by the astronomical data, independent of the accuracy of the reference frequency. Furthermore, these frequencies are derived independently of the [\thCII] centroid, providing an independent verification of the [\thCII] centroid frequency. The resulting centroid frequency (Table~\ref{tab:freq_astro}), derived from astronomical observations, is in good agreement with the value previously determined in the laboratory by \cite{Cooksy1986}.

    \begin{table}[!tb]
        \caption{Astronomically determined [\thCII] transition frequencies and velocity offsets using the [\thCII] centroid and the [\twCII] line as reference frequencies.}
        \label{tab:freq_astro}
        \begin{tabular}{lcc}   
        \hline
        \hline
        \multicolumn{3}{c}{Ref. frequency [\thCII]$^{(1)}$}  \\
        \hline
        Transition line  & Frequency & Velocity offset  \\
         & $f$ & $v_{\mathrm{F' - F}}$  \\
          & [GHz] & [km$\,$s$^{-1}$]  \\
        \hline    
        [$^{13}$\CII] $^2P_{3/2} - ^2P_{1/2}$ &   1900.5458(11) &  \\
        
        [$^{13}$\CII] ${F(2 - 1)}$   & 1900.4648(11) & +11.38 $\pm$ 0.17   \\
        
        [$^{13}$\CII] ${F(1 - 0)}$   & 1900.9511(11) & $-$65.34 $\pm$ 0.17     \\
        
        [$^{13}$\CII] ${F(1 - 1)}$   & 1900.1404(11) & +62.55 $\pm$ 0.17   \\
        \hline
         \multicolumn{3}{c}{Ref. frequency [\twCII]$^{(2)}$}  \\
        \hline
        [$^{13}$\CII] $^2P_{3/2} - ^2P_{1/2}$ &   1900.5459(7) &  \\
        
        [$^{13}$\CII] ${F(2 - 1)}$   & 1900.4649(7) & +11.357 $\pm$ 0.009   \\
        
        [$^{13}$\CII] ${F(1 - 0)}$   & 1900.9512(7) & $-$65.359 $\pm$ 0.024     \\
        
        [$^{13}$\CII] ${F(1 - 1)}$   & 1900.1405(7) & +62.522 $\pm$ 0.017   \\
        \hline
        \end{tabular}
        \tablefoot{(1) The [\thCII] centroid frequency was used as reference determined by \cite{Cooksy1986}.
        (2) The [\twCII] line observed in space with its rest frequency from \cite{Cooksy1986} was used as reference. }
    \end{table}
    
\section{Summary and conclusion}

    In a pioneering study, \cite{Cooksy1986} determined the [\thCII] transition frequencies. However, the laser magnetic resonance method used in that work did not allow for independent measurements of both hfs constants, $A_{1/2}^{\rm hf}$ and $A_{3/2}^{\rm hf}$; the frequency offsets between the three hfs lines therefore had to rely on numerical quantum mechanical calculations to derive the hfs constants. While these calculations have become more precise over time \citep{Joensson1996,Joensson2010}, enabling a more accurate determination of the [\thCII] transition frequencies, no laboratory measurements have been performed to independently validate the computed values.  

    Significant technological advancements in submillimeter observations, such as with the upGREAT receiver \citep{Risacher2016ITTST}, allow for high spectral resolution observations of [\twCII] and all three [\thCII] hyperfine transition lines. The directly measured frequency differences between the hfs lines enabled us to determine both hfs constants with high precision, as summarized in Table~\ref{tab:hfs_astro}.  

    However, without independent knowledge of the systemic velocity of the emitting gas, the absolute rest frequencies of the individual transitions cannot be derived. A reference frequency from laboratory measurements, where the rest frame is well defined, is thus required. In this work, we applied two complementary methods to determine updated [\thCII] transition frequencies:

    \begin{enumerate}
        \item Using the [\thCII] centroid as a reference: By combining the centroid frequency with the hfs constants derived from the astronomical measurements, we determined updated [\thCII] transition frequencies using Eq.~(\ref{eq:hfs_lines}). While the hfs constants themselves can be determined with exceptional accuracy (Table~\ref{tab:hfs_astro}), the uncertainty of the derived transition frequencies is dominated by the uncertainty of the laboratory-measured [\thCII] centroid.
        
        \item Using the frequency shift between [\twCII] and [\thCII] lines from astronomical observations: This method relies on the cleanest [\twCII] spectra with minimal optical depth effects. Its precision is limited by optical depth effects in [\twCII] and the uncertainty of the [\twCII] rest frequency. Nevertheless, it provides an independent determination of the [\thCII] centroid frequency.
    \end{enumerate}

    The derived frequencies from both methods are summarized in Table~\ref{tab:freq_astro}. Both approaches yield consistent frequencies with comparable uncertainties and additionally provide an independent verification of the [\thCII] centroid frequency measured by \cite{Cooksy1986}.

    In this work, we have demonstrated that spectrally resolved astronomical observations can be used to infer atomic or molecular structure, effectively using outer space as a spectroscopic laboratory. The method presented here is not limited to the ionized carbon lines analyzed in this study, but can be applied more broadly to other atoms or molecules where laboratory measurements are challenging. Whenever spectroscopic signatures allow precise measurements of frequency differences (i.e., when two or more lines share the same systemic velocity and are not strongly affected by optical depth effects), this approach leverages the unique synergy between laboratory science and astronomical observations to obtain highly accurate spectroscopic parameters.

\begin{acknowledgements}
    We thank the anonymous referee for their constructive comments, which helped to improve the clarity and quality of this work.
    We gratefully acknowledge support from the Collaborative Research Center 1601 (SFB 1601, sub-projects A6, B2, B7, B8, C1, C2, C3, C4 and C6) and through the Cluster of Excellence ``Our Dynamic Universe''
    under Germany’s Excellence Strategy, both funded by the Deutsche Forschungsgemeinschaft (DFG) – 500700252, and EXC 3037 – 533607693. S.K. acknowledges support by the BMWi via DLR, project number 50OR2311.
\end{acknowledgements}

\bibliographystyle{aa} 
\bibliography{bibliography} 

\clearpage
\appendix

\section{Theoretical updates }\label{sec:theory}

    \begin{table*}[!htb]
        \caption{Literature overview of [$^{13}$\CII] frequencies, based on laboratory measurements combined with theoretical computations. }
        \label{tab:freq_overview}
        \begin{tabular}{lccccc}
        \hline
        \hline
         & \cite{Ossenkopf2013} & \multicolumn{2}{c}{\cite{Cooksy1986}$^{(1)}$} & \multicolumn{2}{c}{\cite{Kramida_and_Haris2022}$^{(2)}$} \\
        \hline
        Transition line & Relative intensity & Frequency & Velocity offset & Frequency & Velocity offset \\
         & $s_{F' - F}$ & $f^{(3)}$ & $v_{F' - F}$ & $f^{(3)}$ & $v_{F' - F}$ \\
         & & [GHz] & [km$\,$s$^{-1}$] & [GHz] & [km$\,$s$^{-1}$] \\
        \hline
        [$^{12}$\CII] $^2P_{3/2} - ^2P_{1/2}$ &  & 1900.5369(7) &    \\
    
        [$^{13}$\CII] $^2P_{3/2} - ^2P_{1/2}$ &  & 1900.5458(11) &    \\
    
        [$^{13}$\CII] $F(2 - 1)$ & 5/8 & 1900.4661(12) & +11.17 $\pm$ 0.18 & 1900.4654(12) & +11.28 $\pm$ 0.19   \\
    
        [$^{13}$\CII] $F(1 - 0)$ & 2/8 & 1900.9500(75) & $-$65.2 $\pm$ 1.2 & 1900.9499(18) & $-$65.1 $\pm$ 0.3   \\
    
        [$^{13}$\CII] $F(1 - 1)$ & 1/8 & 1900.1360(50) & +63.2 $\pm$ 0.8 & 1900.1398(14) & +62.64 $\pm$ 0.21   \\
        \hline
        \end{tabular}
        \tablefoot{(1)~The errors in \cite{Cooksy1986} are given as 2$\sigma$, while in this paper, the errors are represented at the $1\sigma$ level.
        (2)~The frequencies are obtained by combining the hfs constants reported by \cite{Kramida_and_Haris2022} with the centroid frequency determined by \cite{Cooksy1986}.    
         (3)~The numbers in parentheses give the $1\sigma$ uncertainty in the last digits of the quoted frequencies.}
    
    \end{table*}

    \begin{figure}[htb]
        \centering
        \includegraphics[width=.99\columnwidth]{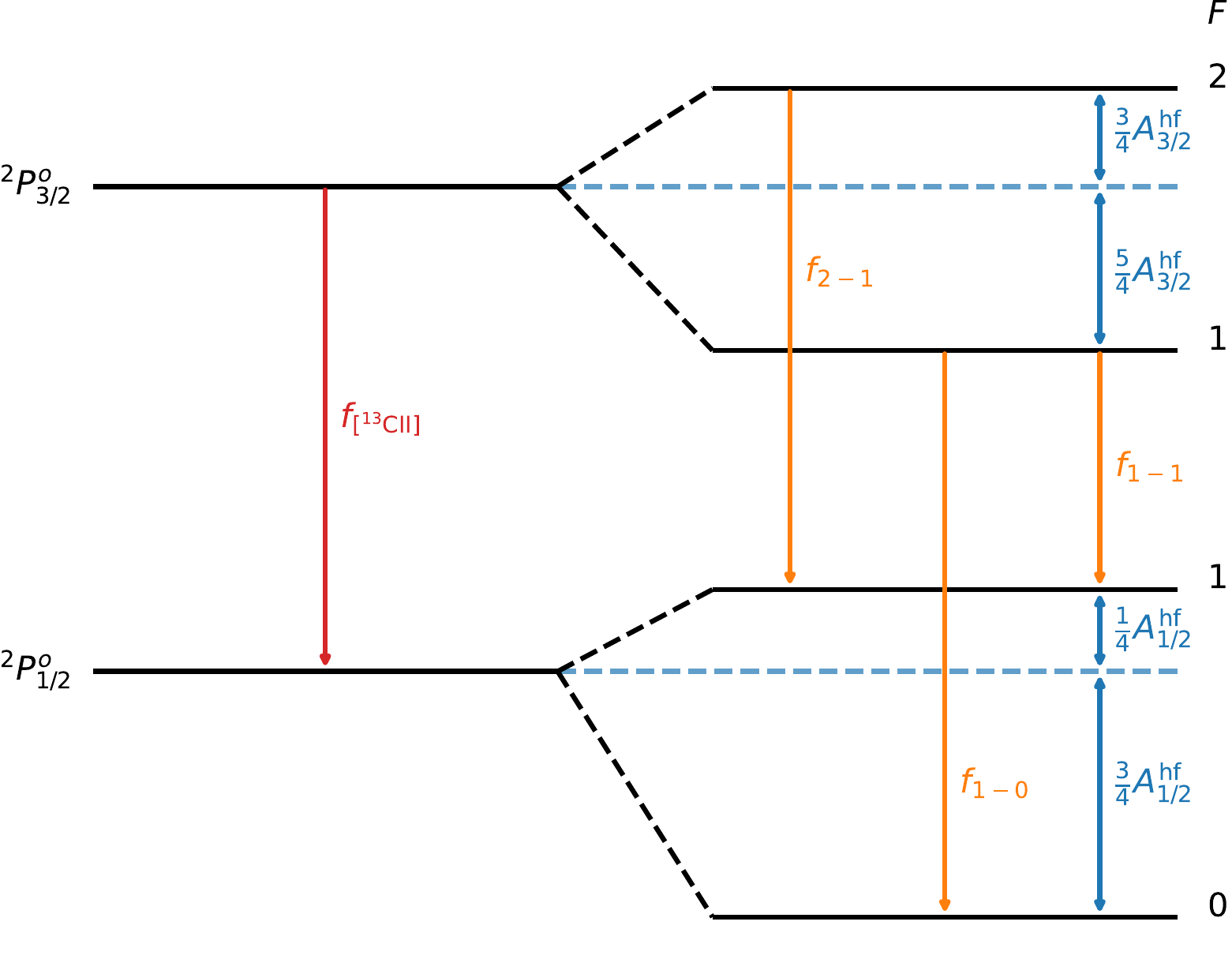}
        \caption{[\thCII] hyperfine-structure energy-level scheme. The red arrow marks the [\thCII] line centroid, the orange arrows indicate the individual hyperfine transitions, and the blue arrows denote the frequency shifts induced by the hfs constants.
        }
        \label{fig:13cii_energies_intro}
    \end{figure}
    
    A schematic representation of the [\thCII] energy levels is shown in Fig.~\ref{fig:13cii_energies_intro}. 
    The two fine-structure levels are denoted $^2P_J$, where $J$ is the total angular momentum quantum number of the unpaired electron, taking the values $J = 1/2$ and $J = 3/2$. 
    The nuclear spin, $I=1/2$, of the $^{13}$C nucleus gives rise to hyperfine splitting, with the hyperfine levels designated by the total angular momentum quantum number $F=J+I$. 
    The splittings of the levels are given by the hyperfine constants and amount to $2A_{3/2}^{\rm hf}$ and $A_{1/2}^{\rm hf}$, respectively (see blue arrows in Fig.~\ref{fig:13cii_energies_intro}). 
    The three hyperfine transitions  (indicated by orange arrows) arise from the selection rule $\Delta F = 0, \pm1$.
    
    Updated calculations of the [\thCII] hyperfine constants have improved since \cite{Cooksy1986}, allowing a more precise determination of the transition frequencies. By anchoring these calculations to the [\thCII] centroid frequency measured in the laboratory, the accuracy of all three hyperfine transition frequencies can be substantially enhanced. \cite{Kramida_and_Haris2022} compiled all available data on the level energies of the C$^+$ isotopes. For the levels treated in the two large-scale calculations of \cite{Joensson1996, Joensson2010}, they adopted weighted averages of the corresponding $A^{\rm hf}$ values. For the ground levels considered in this work, these averages were then used as input to a least-squares optimization procedure constrained by the combined magnetic-dipole hyperfine-structure constant $(A_{1/2}^{\rm hf}-3A_{3/2}^{\rm hf})/4$ measured in the laboratory by \cite{Cooksy1986}. Therefore, the $A_{1/2}^{\rm hf}$ and $A_{3/2}^{\rm hf}$ values reported by \cite{Kramida_and_Haris2022} are semi-empirical.
    
    However, they do not directly list the transition frequencies or level energies but energy differences, $\Delta E_{1-0}$ and $\Delta E_{2-1}$, alongside the corresponding magnetic-dipole hfs constants. We can use these updated calculations to estimate the revised frequencies. The frequency difference between the $F=2$ and $F=1$ levels in the $^2P_{3/2}$ state corresponds to $\Delta E_{2-1}/h=325.6 \pm 1.3$~MHz and the difference between the $F=1$ and $F=0$ levels in the $^2P_{1/2}$ state is $\Delta E_{1-0}/h=810.1 \pm 1.6$~MHz. The individual frequencies then follow from the centroid frequency shifted by a combination of the hfs constants weighted with their relative strength: 
    \begin{eqnarray}
        f_{2-1} &=& f_{\rm [^{13}CII]} - s_{1-0}A_{1/2}^{\rm hf} + (1-s_{2-1})\times 2 A_{3/2}^{\rm hf}\nonumber\\ &=& f_{\rm [^{13}CII]} - \frac{1}{4}A_{1/2}^{\rm hf} + \frac{3}{4}A_{3/2}^{\rm hf}\nonumber\\[4pt]
        f_{1-0} &=& f_{\rm [^{13}CII]} + (1-s_{1-0})A_{1/2}^{\rm hf} - s_{2-1}\times 2 A_{3/2}^{\rm hf}\nonumber\\ &=& f_{\rm [^{13}CII]} + \frac{3}{4}A_{1/2}^{\rm hf}- \frac{5}{4}A_{3/2}^{\rm hf} \nonumber\\[4pt]
        f_{1-1} &=& f_{\rm [^{13}CII]} - s_{1-0}A_{1/2}^{\rm hf} - s_{2-1}\times 2 A_{3/2}^{\rm hf}\nonumber\\ &=& f_{\rm [^{13}CII]} - \frac{1}{4}A_{1/2}^{\rm hf}-\frac{5}{4}A_{3/2}^{\rm hf}~.
        \label{eq:hfs_lines}
    \end{eqnarray}
    The values thus derived for the three transition frequencies and the corresponding velocity offsets relative to the [\twCII] line are given in Table~\ref{tab:freq_overview}. The uncertainties quoted there are obtained through error propagation.

    When discussing error propagation, two types of uncertainties must be considered: the relative frequency uncertainties between the individual [\thCII] hyperfine transitions, and the absolute frequency uncertainty, which includes the uncertainty of the reference frequency. The former is important for correctly superimposing the three hyperfine lines in astronomical observations in order to determine the absolute transition strengths and the intrinsic emission profile. These relative frequency offsets can be derived independently of any reference frequency. However, when estimating the absolute frequencies of the three [\thCII] transitions, the uncertainty of the reference frequency must also be taken into account.

\section{Data reduction}\label{sec:data_reduction}

    The observations were performed using the hexagonal $2\times 7$ pixel low-frequency array (LFA), in horizontal (LFAH) and vertical (LFAV) polarization. The effective size of the dish is $2.5\,\rm{m}$ resulting in a beam size of $14.1$ arcseconds \citep{Risacher2018}. 
    For this study, we examined the SOFIA archive for long-integration [\CII] observations that could be used for our analysis. However, only four observations, spanning three different regions in Orion, provided sufficiently high S/N in all three [\thCII] lines with low contamination of the $F=2-1$ transition: (i) The Orion Bar was observed at its peak emission using a single pointing, covering the emission peak and six additional positions around it with the hexagonal array. These observations were obtained as part of the square-degree [\CII] mapping project (PI: A.G.G.M.~Tielens); see \citet{Higgins2021}.  
    (ii) NGC~1977 was observed with a single pointing toward its emission peak (PI: C.~Guevara), and the entire bright rim of NGC~1977 was mapped in [\twCII] and [\thCII] (PI: S.~Kabanovic). Although the two datasets could be combined, we keep them separate to provide an internal consistency check of our results. (iii) NGC~2024 was mapped in [\twCII] and [\thCII] with long integration times (PI: U.U.~Graf), providing the third source for our analysis.
    
    Each observation was resampled to a spectral resolution of $0.1\,\rm{km\,s^{-1}}$ ($\sim 0.6\,{\rm MHz}$) to reduce the noise level while preserving sufficiently high spectral resolution to obtain an accurate fit to the hfs line positions.
    For the single-pointing observations, all pixels of the array were averaged to produce a single high S/N spectrum. For the spectral cubes, the data were convolved to a spatial resolution of $60''$ and resampled onto a grid with a pixel size of $20''$. A subsequent Gaussian fit was performed for each spectrum along the grid on the two outer [\thCII] lines to determine whether [\thCII] emission is detected (above $3\,\sigma$ for all [\thCII] lines). All grid points with a positive detection were then averaged to produce a single spectrum. 
    
    Spectral baselines in astronomical observations are often affected by standing waves, offsets, or power drifts between calibrations, which must be corrected prior to any analysis. A common approach is to mask the line emission and fit an $m$-order (typically third-order) polynomial to the remaining baseline. The fitted polynomial is then subtracted from the entire spectrum. While this method works well in many cases, using a fixed polynomial order has the disadvantage that complex baselines may not be properly fitted. Residual baseline artifacts remain in some spectra, potentially affecting subsequent analyses. In severe cases, the affected spectra have to be discarded altogether, leading to a loss of valuable observing time. Although higher-order ($m>3$) polynomials can improve the fit, they carry the risk of over-fitting (i.e., by introducing artificial features into the baseline and the emission line).

    To overcome this problem, we apply an adaptive polynomial fitting procedure to each spectrum, minimizing the Corrected Akaike Information Criterion (AICc):
    
    \begin{equation}
        {\rm AICc}  =n\ln(\sigma^2) + 2k + \frac{2k^2 + 2k}{n - k - 1} = {\rm AIC} + \frac{2k^2 + 2k}{n - k - 1} ~,
        \label{eq:aic}
    \end{equation}
    
    \noindent
    where $n$ is the number of fitted channels, $\sigma^2$ is the variance of the baseline, and $k$ is the number of free fit parameters. The last term of the AICc expression is a penalty for higher polynomial orders for fits involving a small number of channels, thereby avoiding over-fitting. We begin with a third-order polynomial fit and subsequently test whether higher-order polynomials provide a statistically better fit according to the AICc. 
    
    If single-channel spikes are present in the spectra, they can be removed by replacing the affected channel values with a linear interpolation between the two neighboring channels. However, such spikes were only found to affect channels outside the signal region. In addition, each scan was inspected visually prior to the averaging or gridding process to remove any potentially bad spectra.
    
\section{Multi-Gaussian fit of the [\thCII] lines}\label{sec:gauss_fit}
    
    \begin{table*}[t]
        \centering
        \caption{Velocity and frequency shift between the [$^{13}$\CII] hyperfine transition lines.}
        \label{tab:13cii_hft_shift}
        \begin{tabular}{l l l@{ $\pm$ }l l l@{ $\pm$ }l l l}
        \hline\hline
        \multicolumn{1}{c}{Source} & \multicolumn{1}{c}{$\Delta f_{2-1}^{(1)}$} & \multicolumn{2}{c}{$\Delta v_{2-1}^{(2)}$} & \multicolumn{1}{c}{$\Delta f_{1-0}^{(1)}$} & \multicolumn{2}{c}{$\Delta v_{1-0}^{(2)}$} & \multicolumn{1}{c}{$A^{\rm hf\, (3)}_{1/2}$} & \multicolumn{1}{c}{$A^{\rm hf\,(3)}_{3/2}$} \\
        & [MHz] & \multicolumn{2}{c}{[km$\,$s$^{-1}$]} & [MHz] & \multicolumn{2}{c}{[km$\,$s$^{-1}$]} & [MHz] & [MHz] \\
        \hline
        NGC 2024 & $324.30(22)$ & $-51.16$ & $0.03$ & $810.73(25)$ & $-127.89$ & $0.04$ & $810.73(25)$ & $162.15(11)$ \\
        Orion Bar & $324.4(3)$ & $-51.17$ & $0.05$ & $810.8(3)$ & $-127.90$ & $0.05$ & $810.8(3)$ & $162.19(14)$ \\
        NGC 1977 Rim & $324.4(3)$ & $-51.17$ & $0.04$ & $810.5(3)$ & $-127.86$ & $0.05$ & $810.5(3)$ & $162.21(13)$ \\
        NGC1977 Peak & $324.37(12)$ & $-51.167$ & $0.019$ & $810.71(14)$ & $-127.883$ & $0.022$ & $810.71(14)$ & $162.19(6)$ \\
        \hline
        Combined observations & $324.37(9)$ & $-51.166$ & $0.015$ & $810.71(11)$ & $-127.882$ & $0.017$ & $810.71(11)$ & $162.18(5)$ \\
        \hline
        \end{tabular}
        \tablefoot{(1) Frequency shift between the [$^{13}$\CII] hyperfine transitions $F = 2 - 1$ and $F=1 - 0$ relative to the $F = 1 - 1$ line, thus $\Delta f_{2-1} = f_{2-1}-f_{1-1}$ and $\Delta f_{1-0} = f_{1-0}-f_{1-1}$.
        (2) Corresponding velocity shift $\Delta v_{2-1} = v_{2-1}-v_{1-1}$ and $\Delta v_{1-0} = v_{1-0}-v_{1-1}$.
        (3) Corresponding hfs constants.}
    \end{table*}

    The spectrally resolved observations of all three hyperfine [\thCII] lines allow us to directly determine the magnetic-dipole hfs constants $A_{J}^{\rm hf}$ by measuring the velocity differences between the lines. To do so, we perform a multi-Gaussian fit to all three lines assuming that the emission is optically thin and can therefore be well represented by a superposition of $N$ Gaussian components:
    
    \begin{equation}
        \begin{aligned}
        T_{\rm mb} &= \sum_{i=1}^{N} T_{i,{\rm [^{13}CII]}} 
        \Bigg[ 
            \sum_{F'-F} s_{F'- F} \,
                \exp\Big(-\beta \Big(\frac{v - v_{i,0} - v_{F'-F}}{w_i}\Big)^2 \Big) \\
        &\quad + \; \alpha \,
                \exp\Big(-\beta \Big(\frac{v - v_{i,0}}{w_i}\Big)^2 \Big)
        \Bigg]~.
        \end{aligned}
        \label{eq:hfs_fit}
    \end{equation}
    \noindent
    The first sum accounts for multiple Gaussian velocity components that may be present in the spectra, with $T_{i,{\rm [^{13}CII]}}$ denoting the total peak [\thCII] temperature of each Gaussian component. Thanks to the high spectral resolution and S/N, even the weak [\thCII] lines allow multiple components to be reliably distinguished. The number of velocity components was determined by minimizing the Bayesian Information Criterion,
        \begin{equation}
            \mathrm{BIC} = \chi^2 + k \ln(n),
        \end{equation}
    where $\chi^2 = \sum_i (T_{\rm mb}(v_i) - T_{\rm fit}(v_i))^2 / \sigma_{\rm rms}^2$ is the chi-squared, $T_{\rm mb}(v_i)$ and $T_{\rm fit}(v_i)$ are the observed main beam temperature and the multi-Gaussian fit at velocity channel $v_i$, $\sigma_{\rm rms}$ is the root-mean-square noise of the spectrum, $k$ is the number of free parameters, and $n$ the number of data points.

    The second sum runs over the three hyperfine transitions ${F' - F} = 2-1,1-0,1-1$, where the total [\thCII] peak intensity of each velocity component is multiplied by the relative strength of each hyperfine line $s_{F'-F}$. The parameter $v_{i,0}$ is the systemic velocity of the component, and $v_{F'-F}$ is the velocity shift relative to the [\twCII] line. The factor $\beta = 4\ln2$ converts the line widths $w_i$ from the standard deviation $\sigma_i$ to the full width at half maximum (FWHM).
    
    Although we selected sources where the strongest hfs line is well separated from the [\twCII] emission, a small level of contamination of the $F=2-1$ line cannot be excluded. To account for a possible contribution from the [\twCII] wing in a self-consistent manner, we scale the Gaussian components using the local carbon isotope abundance ratio, $\alpha = {^{12}{\rm C}}/{^{13}{\rm C}} = 67 \pm 3$ \citep{Langer1990}, assuming that the [\twCII] wing emission is optically thin. We mask the bulk emission of the [\twCII] line, which may be affected by optical depth effects, and fit only the [\twCII] line wings together with the [\thCII] lines to account for any potential contamination. 
    
\begin{figure*}[htb]
   
        \centering
        \includegraphics[width=1.\textwidth]{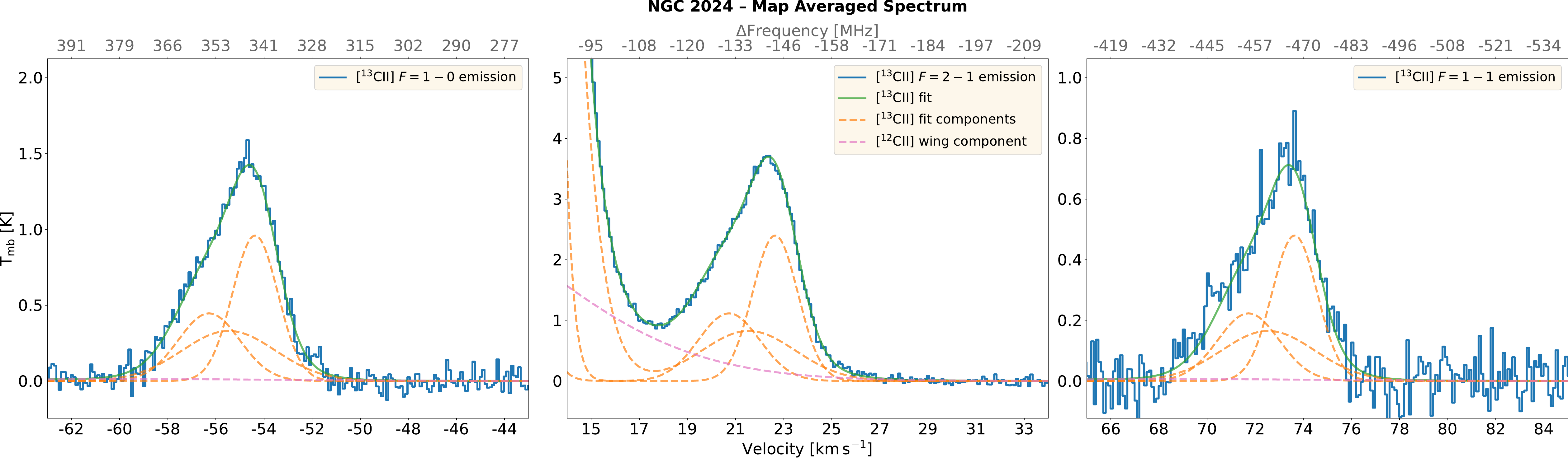}\\
        \includegraphics[width=1.\textwidth]{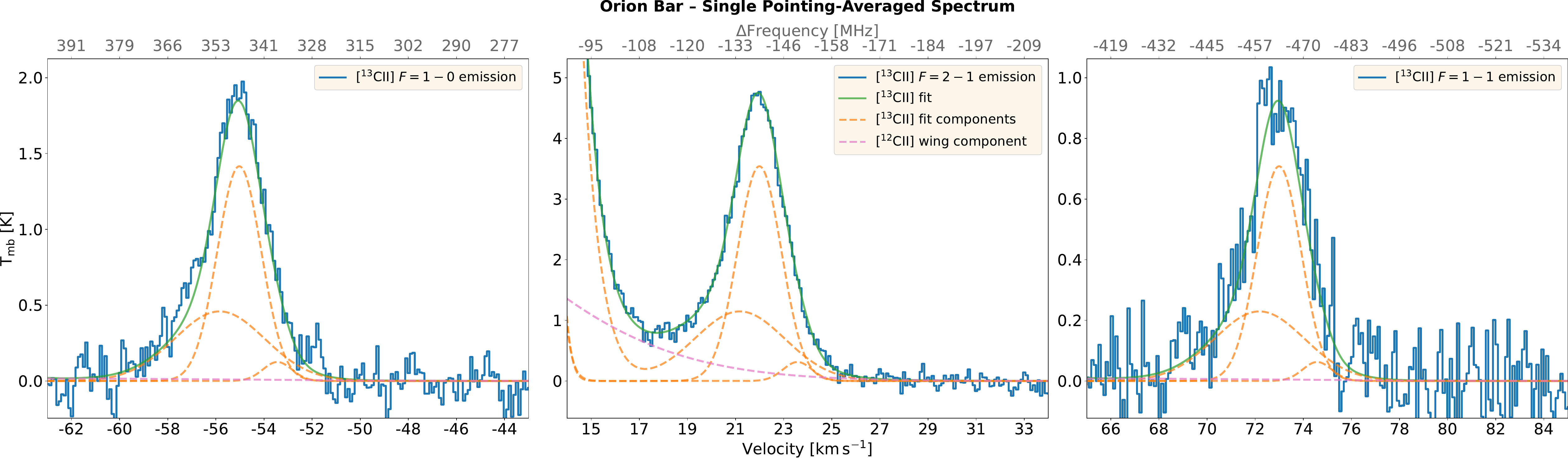}\\
        \includegraphics[width=1.\textwidth]{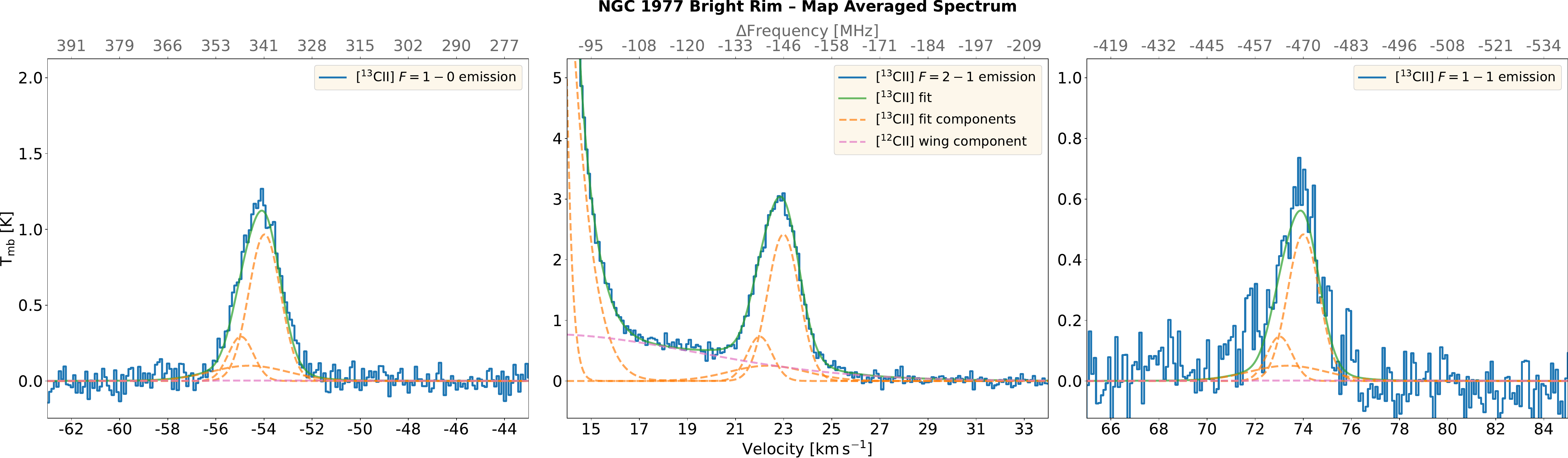}\\
        \includegraphics[width=1.\textwidth]{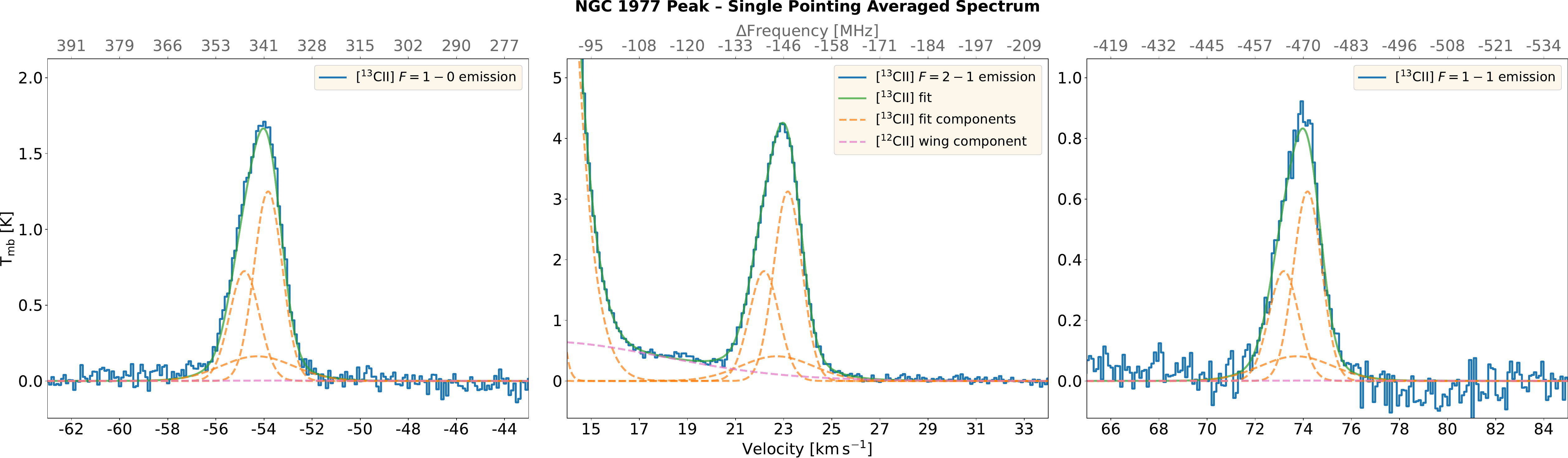}\\
        \caption{Gaussian fits to the three [\thCII] hyperfine structure lines observed toward four sources. From top to bottom: averaged spectrum of the NGC~2024 map, averaged spectrum of the single-pointing observation toward the Orion Bar, averaged spectrum of the bright rim in the NGC~1977 map, and averaged spectrum of the single-pointing observation toward NGC~1977. From left to right: [\thCII] $F=1-0$, $F=2-1$, and $F=1-1$ emission. The blue step curve shows the observed spectrum, the green solid line the total model fit, and the orange dashed curves the individual [\thCII] Gaussian components. The pink dashed curve indicates the [\twCII] wing component, which contributes to the total fit but whose corresponding [\thCII] emission is too faint to be visible.}
        \label{fig:hfs_gauss_fit}
  
    \end{figure*}

    The fit results for each of the four averaged spectra are presented in Fig.~\ref{fig:hfs_gauss_fit} and the corresponding fit parameters are given in Tables~\ref{tab:gauss_fit_ngc2024_map}, \ref{tab:gauss_fit_orion_bar}, \ref{tab:gauss_fit_ngc1977_map}, and \ref{tab:gauss_fit_ngc1977_sp}. The individual Gaussian components contributing to the observed [\thCII] hfs lines are shown as orange dashed curves in Fig.~\ref{fig:hfs_gauss_fit} and are denoted as [\thCII] components.
    In addition to these narrow components, all [\twCII] lines exhibit a broad wing component that is not detected in the [\thCII] spectra due to its low amplitude (when scaled by the local carbon abundance ratio, the corresponding [\thCII] emission is buried below the noise floor). This broad component is represented by the dashed pink line in Fig.~\ref{fig:hfs_gauss_fit} and is denoted as the [\twCII] wing component. This component is not treated any differently from the other components by the model fit. We simply visualize it in a different color because it contributes to the broad [\twCII] emission, which slightly overlaps with the strongest [\thCII] $F=2-1$ emission, introducing a small degree of contamination. This contamination is modeled in all spectra as one broad Gaussian component (the Orion Bar additionally includes multiple weak blue-shifted Gaussian components, which do not contaminate or contribute meaningfully to the [\thCII] lines; these are also labeled as wing components in Table~\ref{tab:gauss_fit_orion_bar}). Such contamination could potentially affect the measured frequency difference between the strongest and weakest hyperfine components, $\Delta v_{2-1} = v_{2-1}-v_{1-1}$. To evaluate this effect, we performed additional fits excluding the broad wing component. The resulting frequency differences changed only marginally, remaining well within the fit uncertainties.

    The results of the hfs fitting for each individual source, together with those for the combined dataset, are summarized in Table~\ref{tab:13cii_hft_shift}. The individual observations are combined using a weighted mean, where the weights are given by $1/\sigma^2$. The table lists the velocity and frequency offsets of the $F=2-1$ and $F=1-0$ components relative to the $F=1-1$ transition, along with the derived hfs constants.
    
    The uncertainties on the derived velocity separations of the [\thCII] hyperfine lines are obtained from the diagonal elements of the covariance matrix returned by the non-linear least-squares fit. The spectral noise (rms), measured from emission-free channels of the averaged spectrum, serves as the per-channel uncertainty, ensuring that the covariance matrix yields properly scaled $1\sigma$ statistical uncertainties. A reduced chi-squared value of $\chi^2_{\rm red} \approx 1.0$ for all fitted spectra confirms that the model adequately describes the data. All reported values and their uncertainties are rounded following the ``rule-of-$m$'' convention with $m = 25$, as described in \cite{Kramida2011}.

    Beyond the formal statistical uncertainties, several potential sources of systematic error were carefully considered. First, despite a dedicated baseline treatment utilizing adaptive polynomial fitting (Appendix~\ref{sec:data_reduction}), small residual baseline undulations cannot be entirely excluded, particularly across the relatively wide velocity window (of $40\,{\rm km/s}$) encompassing the [\twCII] line and the [\thCII] $F = 2-1$ component, which is masked out during the baseline fitting. Second, the $F = 2-1$ hyperfine component is partially blended with the [\twCII] emission, which could bias the fitted line separation. To mitigate this, the [\twCII] line and all [\thCII] hyperfine components were fitted simultaneously, explicitly accounting for the contamination in the model. While these measures were taken to minimize systematic effects, we cannot fully rule out systematic contributions. However, the consistency of the derived hyperfine separations across independent observations suggests that residual systematic effects do not dominate the error budget.
    
    The exceptionally high quality of the data from the long-term single-pointing observation toward the peak emission in NGC~1977 dominates the overall statistics. As a result, the other sources have little impact on the combined results and primarily serve as a consistency check. Although the sample size is relatively small, the integration time per source is exceptionally high, particularly toward NGC~1977.
    \clearpage

    \begin{table}[!tb]
        \caption{Gaussian fit parameters of the map averaged spectrum toward NGC~2024.}
        \label{tab:gauss_fit_ngc2024_map}
        \begin{tabular}{lccc}
        \hline
        \hline
        Component & $T_{i,{\rm [^{13}CII]}} $ & $v_{i,0}$ & $w_i$ \\
                  & [K] & [km\,s$^{-1}$] & [km\,s$^{-1}$] \\
        \hline
        [\thCII] comp.\,1 & $1.8 \pm 0.4$ & $9.37 \pm 0.24$ & $3.0 \pm 0.4$ \\
    
        [\thCII] comp.\,2 & $3.8 \pm 0.4$ & $11.29 \pm 0.05$ & $2.21 \pm 0.07$ \\
    
        [\thCII] comp.\,3 & $1.3 \pm 0.3$ & $10.19 \pm 0.22$ & $4.59 \pm 0.09$ \\
    
        [\twCII] wing & $0.048 \pm 0.008$ & $6.54 \pm 0.22$ & $14.98 \pm 0.08$ \\
    
        \hline
        \end{tabular}
        \tablefoot{The reduced $\chi^2$ of the fit is 1.16. The rms noise of the spectrum is 0.058\,K. The S/N of the weakest [\thCII] $F=1$--$1$ hyperfine component is 12.2.}
    \end{table}

    \begin{table}[!tb]
        \caption{Gaussian fit parameters of the single-pointing averaged spectrum toward the Orion Bar.}
        \label{tab:gauss_fit_orion_bar}
        \begin{tabular}{lccc}
        \hline
        \hline
        Component & $T_{i,{\rm [^{13}CII]}} $ & $v_{i,0}$ & $w_i$ \\
                  & [K] & [km\,s$^{-1}$] & [km\,s$^{-1}$] \\
        \hline
        [\thCII] comp.\,1 & $5.7 \pm 0.3$ & $10.63 \pm 0.05$ & $2.15 \pm 0.10$ \\
    
        [\thCII] comp.\,2 & $0.5 \pm 0.3$ & $12.2 \pm 0.3$ & $1.4 \pm 0.4$ \\
    
        [\thCII] comp.\,3 & $1.84 \pm 0.21$ & $9.82 \pm 0.18$ & $4.42 \pm 0.09$ \\
    
        [\twCII] wing\,1 & $0.064 \pm 0.007$ & $3.1 \pm 0.3$ & $16.72 \pm 0.19$ \\
    
        [\twCII] wing\,2 & $0.022 \pm 0.003$ & $2.24 \pm 0.18$ & $1.24 \pm 0.08$ \\
    
        [\twCII] wing\,3 & $0.054 \pm 0.007$ & $-1.42 \pm 0.20$ & $2.38 \pm 0.13$ \\
    
        [\twCII] wing\,4 & $0.029 \pm 0.005$ & $-3.33 \pm 0.23$ & $2.11 \pm 0.15$ \\
    
        [\twCII] wing\,5 & $0.0058 \pm 0.0014$ & $-12.9 \pm 0.4$ & $5.2 \pm 0.8$ \\
    
        [\twCII] wing\,6 & $0.026 \pm 0.003$ & $-19.1 \pm 0.3$ & $9.0 \pm 0.3$ \\
    
        \hline
        \end{tabular}
        \tablefoot{The reduced $\chi^2$ of the fit is 1.00. The rms noise of the spectrum is 0.109\,K. The S/N of the weakest [\thCII] $F=1$--$1$ hyperfine component is 8.4.}
    \end{table}

    \begin{table}[!tb]
        \caption{Gaussian fit parameters of the map averaged spectrum toward NGC~1977.}
        \label{tab:gauss_fit_ngc1977_map}
        \begin{tabular}{lccc}
        \hline
        \hline
        Component & $T_{i,{\rm [^{13}CII]}} $ & $v_{i,0}$ & $w_i$ \\
                  & [K] & [km\,s$^{-1}$] & [km\,s$^{-1}$] \\
        \hline
        [\thCII] comp.\,1 & $3.9 \pm 0.4$ & $11.63 \pm 0.12$ & $1.56 \pm 0.13$ \\
    
        [\thCII] comp.\,2 & $0.40 \pm 0.11$ & $11.0 \pm 0.3$ & $3.76 \pm 0.07$ \\
    
        [\thCII] comp.\,3 & $1.2 \pm 0.7$ & $10.65 \pm 0.20$ & $1.23 \pm 0.22$ \\
    
        [\twCII] wing & $0.012 \pm 0.003$ & $12.4 \pm 0.3$ & $15.0 \pm 0.4$ \\
    
        \hline
        \end{tabular}
        \tablefoot{The reduced $\chi^2$ of the fit is 1.11. The rms noise of the spectrum is 0.074\,K. The S/N of the weakest [\thCII] $F=1$--$1$ hyperfine component is 7.5.}
    \end{table}

    \begin{table}[!tb]
        \caption{Gaussian fit parameters of the single-pointing averaged spectrum toward NGC~1977.}
        \label{tab:gauss_fit_ngc1977_sp}
        \begin{tabular}{lccc}
        \hline
        \hline
        Component & $T_{i,{\rm [^{13}CII]}} $ & $v_{i,0}$ & $w_i$ \\
                  & [K] & [km\,s$^{-1}$] & [km\,s$^{-1}$] \\
        \hline
        [\thCII] comp.\,1 & $5.0 \pm 0.7$ & $11.82 \pm 0.08$ & $1.38 \pm 0.07$ \\
    
        [\thCII] comp.\,2 & $0.65 \pm 0.09$ & $11.36 \pm 0.11$ & $3.55 \pm 0.03$ \\
    
        [\thCII] comp.\,3 & $2.9 \pm 0.6$ & $10.83 \pm 0.15$ & $1.45 \pm 0.14$ \\
    
        [\twCII] wing & $0.0099 \pm 0.0016$ & $12.81 \pm 0.20$ & $12.1 \pm 0.4$ \\
    
        \hline
        \end{tabular}
        \tablefoot{The reduced $\chi^2$ of the fit is 1.03. The rms noise of the spectrum is 0.052\,K. The S/N of the weakest [\thCII] $F=1$--$1$ hyperfine component is 16.0.}
    \end{table}
    
    \clearpage
    
\section{Single-layer model}\label{sec:single_layer_model}

    [\twCII] emission arising from dense and bright PDRs is often affected by optical depth effects. To account for self-absorption, we model the emission by fitting the radiative transfer equation:
    \begin{equation} \label{eq:1layer}
        T_{\mathrm{mb}}(v) = \mathcal{J}_{[\rm{^{12}CII}]}(T_{\rm{ex}}) \, \left( 1-e^{- \sum_i^N\tau_{i}(v)}\right)~.
    \end{equation}
    \noindent
    We assume that the emitting gas can be described by a single excitation temperature, $T_{\rm ex}$. The equivalent brightness temperature is given by:
    \begin{equation}
        \mathcal{J}_{[\rm{^{12}CII}]}(T_{\rm{ex}}) = \frac{T_0}{e^{T_0/T_{\rm{ex}}}-1}~,
    \end{equation}
    \noindent
    where $T_0 = h f_{[\rm{^{12}CII}]}/k_B$ is the equivalent temperature of the transition.

    Equation~(\ref{eq:1layer}) allows for multiple velocity components contributing to the observed spectrum. The optical depth of each component is modeled by a Gaussian velocity profile for each transition line fitted:
    \begin{equation}
        \tau_i(v)= \tau_{i,0}\,e^{-\beta~\left(\frac{v-v_{i,0}}{w_i}\right)^2}+ \sum_{F'-F} \frac{s_{F'-F}}{\alpha}\,\tau_{i,0} \, e^{-\beta~\left(\frac{v-v_{i,0}-v_{F'-F}}{w_i}\right)^2}~,
    \end{equation}
    \noindent
    where $\tau_{i,0}$, $v_{i,0}$, and $w_i$ denote the peak optical depth, systemic velocity, and line width of the $i$-th component, respectively, and $s_{F'-F}$ and $v_{F'-F}$ describe the relative strengths and velocity offsets of the hyperfine transitions. Since the optical depth is given with respect to the [\twCII] emission, the corresponding [\thCII] optical depth is obtained by dividing by the carbon abundance ratio $\alpha$ and multiplying by the relative strength of the corresponding $F' - F$ transition.

    The three [\thCII] hyperfine lines and the [\twCII] line are fitted simultaneously. This approach ensures a self-consistent treatment of the potential contamination of the [\thCII] $F=2-1$ emission by the broad-wing emission of the [\twCII] line.

\begin{figure}[tp]
        \centering
        \includegraphics[width=.99\columnwidth]{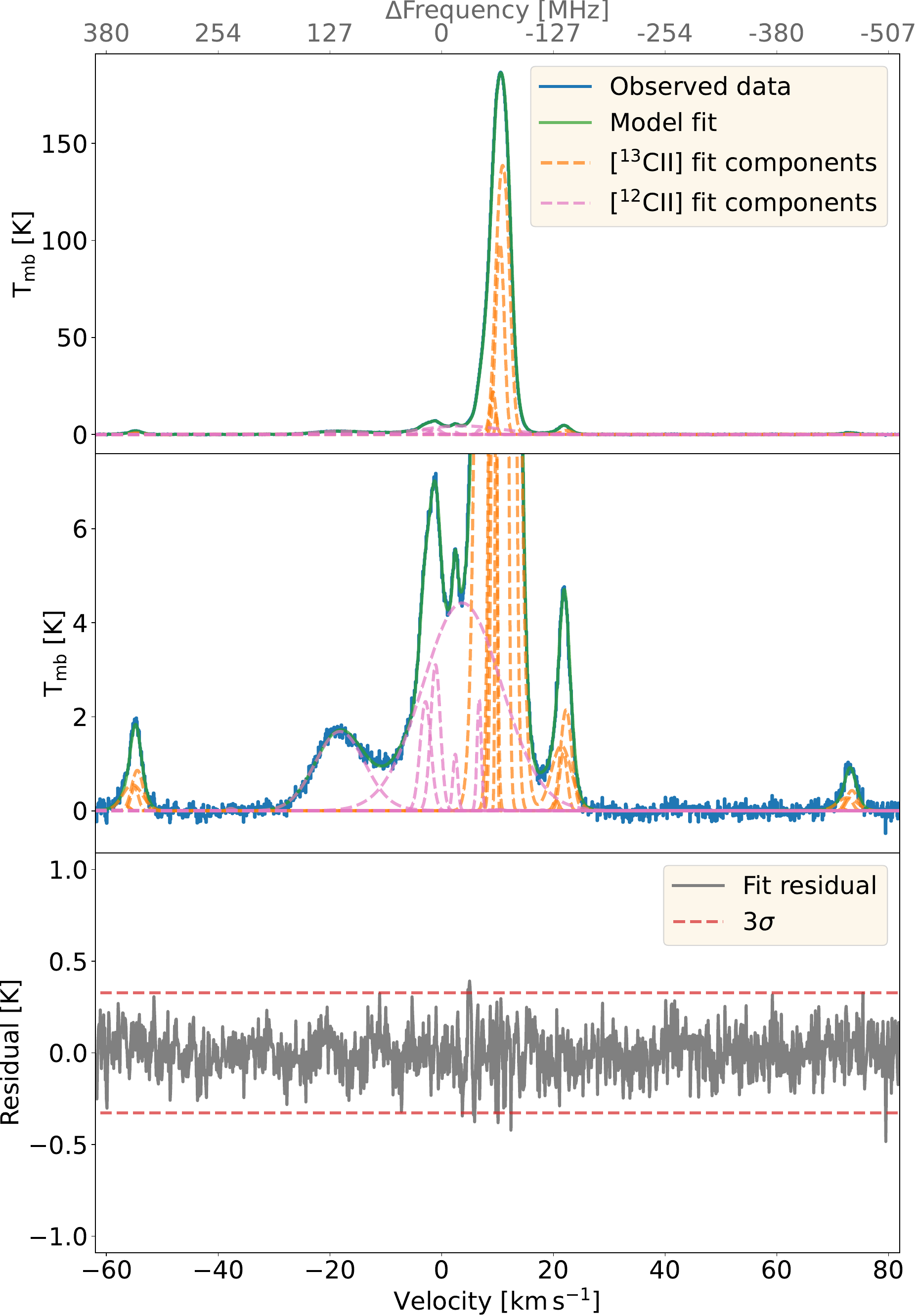}
        \caption{Single-layer model fit. The spectrally resolved [\twCII] and [\thCII] emission from the Orion Bar is shown by the blue data points in the upper panel and in the middle panel, which is zoomed in to highlight the weak [\thCII] emission. The model fit is shown by the green curve. The orange dashed lines indicate the components visible in [\thCII], while the pink components represent the [\twCII] emission that is not visible in [\thCII]. The lower panel shows the residuals between the model and the observed data, and the red horizontal dashed lines indicate the $3\,\sigma$ level.}
        \label{fig:single_layer_model}
    \end{figure}

    Fig.~\ref{fig:single_layer_model} shows the best-fit model obtained by solving the radiative transfer Eq.~(\ref{eq:1layer}), with the corresponding fit parameters given in Table~\ref{tab:single_layer_fit}. The [\twCII] spectrum exhibits multiple velocity components. The strong bulk emission, represented by the dashed orange components, is also detected in the [\thCII] lines. Owing to the higher S/N of the [\twCII] data, additional velocity components are identified that, when scaled down by the local carbon abundance ratio, fall below the noise level and are therefore not visible in the [\thCII] profiles. These components are visualized by the pink dashed lines and are predominantly found on the blueshifted side of the line, alongside the broad component, which creates the redshifted [\twCII] wing and slightly contaminates the [\thCII] $F=2-1$ emission. Although the weak blueshifted [\twCII] components do not contribute to the [\thCII] lines, we nevertheless include them in the model to assess their potential influence. In particular, they may modulate the broad [\twCII] component that fits the redshifted wing of the [\twCII] emission, which could in turn affect the [\thCII] $F=2-1$ transition.

    \begin{table}[!tb]
    \caption{Single-layer fit parameters of the averaged spectrum toward the Orion Bar.}
    \label{tab:single_layer_fit}
    \begin{tabular}{lccc}
    \hline
    \hline
    Component & $\tau_{i,0}$ & $v_{i,0}$ & $w_i$ \\
              &  & [km\,s$^{-1}$] & [km\,s$^{-1}$] \\
    \hline
    Comp.\,1 & $1.13 \pm 0.10$ & $10.92 \pm 0.07$ & $2.34 \pm 0.05$ \\

    Comp.\,2 & $0.65 \pm 0.11$ & $10.446 \pm 0.019$ & $1.59 \pm 0.05$ \\

    Comp.\,3 & $0.729 \pm 0.011$ & $10.065 \pm 0.005$ & $4.208 \pm 0.011$ \\

    Comp.\,4 & $0.11 \pm 0.03$ & $9.19 \pm 0.03$ & $1.12 \pm 0.08$ \\

    Comp.\,5 & $0.056 \pm 0.006$ & $9.754 \pm 0.012$ & $0.42 \pm 0.04$ \\

    Comp.\,6 & $0.02181 \pm 0.00019$ & $3.64 \pm 0.06$ & $16.44 \pm 0.09$ \\

    Comp.\,7 & $0.0117 \pm 0.0007$ & $6.723 \pm 0.021$ & $0.88 \pm 0.06$ \\

    Comp.\,8 & $0.0059 \pm 0.0003$ & $2.446 \pm 0.021$ & $1.02 \pm 0.06$ \\

    Comp.\,9 & $0.0153 \pm 0.0012$ & $-1.10 \pm 0.08$ & $2.04 \pm 0.10$ \\

    Comp.\,10 & $0.0114 \pm 0.0009$ & $-2.88 \pm 0.14$ & $2.28 \pm 0.17$ \\

    Comp.\,11 & $0.00829 \pm 0.00008$ & $-18.20 \pm 0.05$ & $10.41 \pm 0.13$ \\

    Comp.\,12 & $0.00025 \pm 0.00019$ & $-37.7 \pm 0.6$ & $1.5 \pm 1.4$ \\

    \hline
    \end{tabular}
    \tablefoot{The excitation temperature is $T_{\rm ex} = 247.3 \pm 0.4$\,K. The reduced $\chi^2$ of the fit is 1.13.}
\end{table}

\end{document}